%% file: qshap.tex
\documentclass[10pt]{article} 
\usepackage{lmodern}
\usepackage[preprint]{tmlr}

\usepackage[hyphens]{url}
\usepackage{cite} 
\usepackage[hidelinks]{hyperref}
\usepackage{booktabs}
\usepackage{amssymb, amsmath, mathrsfs}
\usepackage{adjustbox}
\usepackage{tabularx}  
\usepackage{array}     

\usepackage{xcolor}
\usepackage{outlines}
\usepackage{multirow}
\usepackage{graphicx}
\usepackage{algorithm}
\usepackage{algpseudocode}
\usepackage{subcaption}
\usepackage{tabularx}
\usepackage{xspace}

\definecolor{customblue}{HTML}{2e47be}
\definecolor{hidden-blue}{RGB}{215,238,249}
\definecolor{hidden-black}{RGB}{20,68,106}
\definecolor{takeaway-blue}{RGB}{218,227,243}
\definecolor{takeaway-title-blue}{RGB}{51,74,133}
\definecolor{authorYellow}{HTML}{ff9100}
\definecolor{authorGreen}{HTML}{59C4B8}
\definecolor{authorBlue}{HTML}{4C9CFF}
\definecolor{authorPurple}{HTML}{AF73FA}
\definecolor{authorPink}{HTML}{f0539b}
\definecolor{authorCyan}{HTML}{548c29}

\input{custom_commands.tex}

\hypersetup{
    colorlinks=true,
    citecolor=customblue,
    linkcolor=customblue,
    urlcolor=customblue 
}

\newcommand{\QGShap}{\textsc{QGShap}\xspace}

\begin{document}
\title{\QGShap: Quantum Acceleration for Faithful GNN Explanations}

\author{
    \name\!\!Haribandhu Jena \email haribandhu.jena@niser.ac.in\\
    \addr School of Computer Sciences\\
    National Institute of Science Education and Research\\
    An OCC of Homi Bhabha National Institute, India\\\\
    \AND
    \name Jyotirmaya Shivottam \email jyotirmaya.shivottam@niser.ac.in\\
    \addr School of Computer Sciences\\
    National Institute of Science Education and Research\\
    An OCC of Homi Bhabha National Institute, India\\\\
    \AND
    \name Subhankar Mishra \email smishra@niser.ac.in\\
    \addr School of Computer Sciences\\
    National Institute of Science Education and Research\\
    An OCC of Homi Bhabha National Institute, India
}

\maketitle              

\input{sections/0_abstract}
\input{sections/1_introduction}
\input{sections/2_related_work}
\input{sections/2_background}
\input{sections/3_methodology}
\input{sections/4_experiments}
\input{sections/5_results}

\input{sections/6_conclusion}


\bibliographystyle{tmlr}
\bibliography{qshap}

\end{document}

%% file: custom_commands.tex
\usepackage{xspace}

\newcommand{\fidp}{\mathsf{Fid}^+}
\newcommand{\fidm}{\mathsf{Fid}^-}

\newcommand{\ita}[1]{\textit{#1}}
\newcommand{\bfa}[1]{\textbf{#1}}

\newcommand{\ol}[1]{ 
    \begin{outline}[enumerate]
    #1
    \end{outline}
}
\newcommand{\ul}[1]{ 
    \begin{outline}
    #1
    \end{outline}
}

\definecolor{brightorange}{RGB}{255,127,0} 

\definecolor{tdbg}{HTML}{FFCC00}
\definecolor{fixbg}{HTML}{CC0000}
\definecolor{revbg}{HTML}{EB0CB7}
\definecolor{checkbg}{HTML}{95FA2F}
\definecolor{changebg}{HTML}{12C915}
\definecolor{ongoingbg}{HTML}{19FFA5}
\definecolor{ablatebg}{HTML}{1F3FF5}
\definecolor{modbg}{HTML}{CC00CC}
\definecolor{tdlbg}{HTML}{BFFFF0}
\definecolor{tdnbg}{HTML}{FFE3BF}
\definecolor{notebg}{HTML}{FF0000}
\definecolor{donebg}{HTML}{00C3FF}
\definecolor{textdark}{HTML}{1F1F1F}

\algnewcommand{\Input}[1]{\State \textbf{Input:} #1}
\algnewcommand{\Output}[1]{\State \textbf{Output:} #1}

\newcommand{\best}[1]{\mathbf{#1}}      
\newcommand{\sbest}[1]{\underline{#1}}



%% file: sections/0_abstract.tex
\begin{abstract}
    Graph Neural Networks (GNNs) have become indispensable in critical domains such as drug discovery, social network analysis, and recommendation systems, yet their black-box nature hinders deployment in scenarios requiring transparency and accountability. While Shapley value-based methods offer mathematically principled explanations by quantifying each component's contribution to predictions, computing exact values requires evaluating \(2^n\) coalitions (or aggregating over \(n!\) permutations), which is intractable for real-world graphs. Existing approximation strategies sacrifice either fidelity or efficiency, limiting their practical utility. We introduce \QGShap\footnote{To appear in the CCIS series (Springer Nature), QC+AI Workshop at AAAI 2026.}, a quantum computing approach that leverages amplitude amplification to achieve quadratic speedups in coalition evaluation while maintaining exact Shapley computation. Unlike classical sampling or surrogate methods, our approach provides fully faithful explanations without approximation trade-offs for tractable graph sizes. We conduct empirical evaluations on synthetic graph datasets, demonstrating that \QGShap achieves consistently high fidelity and explanation accuracy, matching or exceeding the performance of classical methods across all evaluation metrics. These results collectively demonstrate that \QGShap not only preserves exact Shapley faithfulness but also delivers interpretable, stable, and structurally consistent explanations that align with the underlying graph reasoning of GNNs. The implementation of \QGShap is available at \url{https://github.com/smlab-niser/qgshap}.
\end{abstract}

%% file: sections/1_introduction.tex
\section{Introduction}

Graph neural networks (GNNs) have gained widespread use for learning from graph-structured data in critical applications such as molecular chemistry~\citep{gnnchem1}, social network analysis~\citep{gnnsocial}, and recommendation systems~\citep{gnnrecom}. They excel at capturing complex relationships and patterns within interconnected data \citep{gnnsurv}, enabling breakthroughs in areas like drug discovery, social network analysis, and recommendation systems. However, despite their success, GNNs often function as `black boxes', making it difficult for users and stakeholders to understand how they arrive at their decisions~\citep{xgnnrev}. This opacity poses significant challenges in domains, where transparency, trust, and accountability are essential. Additionally, the complexity of GNN architectures~\citep{mpnn,gcn,gat} and the diversity of graph data further complicate efforts to interpret their predictions.

Building on recent advances in GNN explainability, researchers have moved beyond node and edge-level explanation methods, such as GNNExplainer~\citep{gnnexp}, PGExplainer~\citep{pge}, and GraphLIME~\citep{glime}, toward approaches that capture more complex structural patterns in graphs. While GNNExplainer and PGExplainer use gradient and perturbation-based techniques to identify important components, they often produce explanations that are faithful but unstable, especially on complex benchmarks~\citep{xgnnrev}. Surrogate learning methods like GraphLIME aim to explain local feature relationships, but they still explain only the node features. More recent work, such as SubgraphX~\citep{subg}, leverages Shapley values~\citep{shapv} combined with Monte Carlo Tree Search~\citep{mcts} to identify entire explanatory subgraphs. This shift enables explanations that are both more faithful and interpretable at a higher semantic level, moving the field toward more robust and meaningful forms of interpretability.

Motivated by this shift toward subgraph-level reasoning, Shapley value-based \citep{shap} explainability methods have emerged as the principled foundation underlying such scoring, offering a mathematically rigorous way to quantify how each node or subgraph contributes to a model's prediction~\citep{subg,gnnshap,graphsvx}. By aggregating the marginal impact of each component across all possible coalitions, Shapley values provide fairness and completeness in attribution~\citep{shap}. Yet, the very strength of this formulation - its exhaustive consideration of all $2^n$ combinations, renders it computationally prohibitive for real-world graphs. Classical approaches have sought to approximate these values through sampling, Monte Carlo estimation, or surrogate modeling, but such strategies unavoidably trade off either fidelity or efficiency~\citep{subg,graphsvx,gnnshap,graphshapiq}.

To move beyond this bottleneck while retaining Shapley's axiomatic benefits, recent advances in quantum computing have introduced algorithms that exploit amplitude amplification to achieve quadratic speedups for combinatorial evaluation tasks, including subset and coalition scoring~\citep{qmcts,qsv}. Building on these developments, we propose \QGShap for GNN explainability that leverages amplitude amplification to accelerate coalition evaluation, while maintaining exact Shapley computation. Empirical evaluations on synthetic and small real-world datasets demonstrate that our method achieves exact Shapley faithfulness, suggesting that quantum algorithms can play a transformative role in scaling the explainability of GNNs.

\paragraph{\bfa{Contributions.}}
\ol{
    \1 We introduce \QGShap, demonstrating that quantum amplitude-estimation techniques can accelerate brute-force Shapley value computation for GNN explanations, achieving faithful, theoretically grounded node attributions with quadratic query speedup over classical approaches.
    \1 \QGShap provides exact Shapley-based explanations and surpasses existing explainers on explanation-quality metrics across synthetic graph benchmarks.
}

%% file: sections/2_related_work.tex
\section{Related Work}

GNNs have established themselves as highly effective frameworks for learning and reasoning over complex structured data~\citep{gnnrev}. Despite their remarkable predictive capabilities and widespread success across numerous domains, the internal decision-making mechanisms of GNNs often remain largely opaque~\citep{xgnnrev}. Early research in GNN explainability focused on attributing importance to individual nodes, edges, or features through gradient-based techniques such as Saliency Analysis (SA)~\citep{sa}, Class Activation Mapping (CAM)~\citep{cam}, and Guided Backpropagation (Guided BP)~\citep{ggp}; decomposition-based methods like Layer-wise Relevance Propagation (LRP)~\citep{lrp} and Excitation Backpropagation (Excitation BP)~\citep{egp}; and perturbation-based approaches that assess model sensitivity to input modifications. Among these, GNNExplainer~\citep{gnnexp} learns soft masks over graph components to maximize the mutual information between original and perturbed predictions, while PGExplainer~\citep{pge} extends this by learning a parameterized probabilistic model that generalizes edge importance prediction across graphs. Surrogate-based frameworks, such as GraphLIME~\citep{glime}, adopt locally interpretable linear models to approximate the neighborhood-level decision boundary of GNNs. These explanation methods are locally accurate but unstable and lack broader, human-friendly explanations~\citep{xgnnrev}.

Recent advances in GNN explainability have centered on Shapley value-based methods~\citep{shapv}, each introducing distinct approximations to make computation tractable for real-world graph data. SubgraphX~\citep{subg} employs Monte Carlo Tree Search~\citep{mcts} and approximates Shapley values through limited sampling, which, while efficient for small graphs, becomes impractically slow for larger or denser graphs due to the exponential coalition space. GraphSVX~\citep{graphsvx} constructs a surrogate model on a perturbed dataset and samples coalitions, but its model-agnostic approach undersamples mid-sized coalitions, potentially reducing explanation fidelity. GNNShap~\citep{gnnshap} leverages GPU parallelism and batching to accelerate Shapley value estimation, achieving significant speedups over prior methods, yet fundamentally remains an approximation technique reliant on sampling rather than exact computation. In contrast, GraphSHAP-IQ~\citep{graphshapiq} exploits the structure of message-passing GNNs with linear global pooling and output layers to compute exact any-order Shapley interactions, scaling near-linearly for sparse and shallow graphs and requiring far fewer model calls than model-agnostic baselines. However, for deep architectures, densely connected graphs, or when the largest $\ell$-hop neighborhood exceeds a practical threshold, GraphSHAP-IQ must introduce a hyperparameter to limit the highest order of computed interactions, trading exactness for tractability. Critically, its theoretical guarantees break down for GNNs with nonlinear readout functions, as interactions extend beyond receptive fields and the sparse interaction property is lost, substantially increasing complexity. As a result, GraphSHAP-IQ cannot be directly applied to quantum GNNs such as Equivariant Quantum Graph Circuits~\citep{eqgc}, where nonlinear quantum interactions and entanglement fundamentally violate the linearity and decomposability assumptions required for the explainability frameworks.

The computation of Shapley values for GNN explanations is intractable in general. Exact computation is \#P-complete for classical (explicit) cooperative games~\citep{deng}, and becomes $\mathsf{FP}^{\#\mathsf{P}}$-hard (and in some cases \#P-complete) in succinct graph or query settings such as RPQs and CRPQs~\citep{icdt,whenshap}. Thus, the exponential coalition enumeration $2^n$ should be viewed as a symptom rather than the formal cause. Consequently, precise evaluation involves considering all $2^n$ possible node subsets for a graph of $n$ nodes in the worst case. This requirement severely limits the scalability of classical explainability methods, as traditional techniques such as sampling and model-agnostic surrogates introduce unavoidable trade-offs between computational efficiency and explanation fidelity. To address the inherent bottlenecks of traditional computation, researchers have increasingly turned to quantum computing as an alternative paradigm~\citep{qae,optq}. Recent work~\citep{qsv} demonstrates a quantum algorithm that encodes coalition weights and marginal contributions into quantum states while employing amplitude amplification to achieve quadratic speedup relative to classical Monte Carlo strategies. By adapting this algorithm to the graph domain, it becomes feasible to compute Shapley values at the subgraph level with all possible coalitions of size $2^n$. This advancement effectively mitigates the exponential coalition bottleneck, thereby enhancing both the scalability and explainability of Shapley-based explanations in graph neural networks.

%% file: sections/2_background.tex
\section{Background}
\subsection{Graph Neural Networks (GNNs)}
Let \(\mathcal{G} = (\mathcal{V}, \mathcal{E}, X)\) denote an undirected graph with node set \(\mathcal{V}\), edge set \(\mathcal{E} \subseteq \mathcal{V} \times \mathcal{V}\), and node feature matrix \(X \in \mathbb{R}^{n \times d}\), where \(n\) is the number of nodes and \(d\) is the dimensionality of node features. A GNN~\citep{mpnn} computes node embeddings by iteratively aggregating neighborhood information:
\begin{equation}
\mathbf{h}_v^{(l)} = U^{(l)}\left(\mathbf{h}_v^{(l-1)},\, \sum_{u \in \mathcal{N}(v)} M^{(l)}\left(\mathbf{h}_v^{(l-1)}, \mathbf{h}_u^{(l-1)}\right)\right)
\end{equation}

where, \(M^{(l)}\) and \(U^{(l)}\) are (possibly learnable) functions, and \(\mathbf{h}_v^{(0)} = \mathbf{x}_v\). For graph-level tasks, node embeddings are aggregated via a readout function \(R\): \(\mathbf{h}_\mathcal{G} = R(\{\!\!\{\mathbf{h}_v^{(L)}\}\!\!\})\).

\subsection{Shapley Values for Explanations}
\label{sec:shapley}

A \emph{coalitional game}~\citep{coalition} is given by \((P, v)\), where \(P = \{1, \ldots, n\}\)
and a function, \(v : 2^{P} \rightarrow \mathbb{R}\) assigns a value to each coalition \(S \subseteq P\), with \(v(\emptyset) = 0\).
For a player \(p_j \in P\), the Shapley value is defined as

\begin{equation}
\phi(p_j) = \sum_{S \subseteq P \setminus \{p_j\}}
w(|S|, n)\,[v(S \cup \{p_j\}) - v(S)]
\end{equation}

where \(w(|S|, n) = \frac{1}{\binom{n-1}{|S|}} \cdot \frac{1}{n}\) is the coalition weighting term.
The quantity \(\phi(p_j)\) is the unique solution that satisfies the following axioms:

\begin{itemize}
    \item \textbf{Efficiency:} The total value is distributed, i.e., \(\sum_{p_j \in P} \phi(p_j) = v(P)\).
    \item \textbf{Symmetry:} If two players \(p_j, p_k\) satisfy \(v(S \cup \{p_j\}) = v(S \cup \{p_k\})\) for all \(S \subseteq P \setminus \{p_j, p_k\}\), then \(\phi(p_j) = \phi(p_k)\).
    \item \textbf{Dummy:} If a player \(p_j\) satisfies \(v(S \cup \{p_j\}) = v(S)\) for all \(S \subseteq P \setminus \{p_j\}\), then \(\phi(p_j) = 0\).
    \item \textbf{Additivity:} For games with value functions \(v\) and \(v'\), the Shapley value for the summed game is \(\phi(p_j)(v + v') = \phi(p_j)(v) + \phi(p_j)(v')\).
\end{itemize}

\subsection{Quantum Estimation of the Shapley Value}
\label{sec:quantum_shapley}

Consider the classical coalitional game $(P, v)$ with $n$ players.  
In the quantum algorithm setup~\citep{qsv}, represent the game as $(P, U)$, where $U: 2^P \rightarrow [0,1]$ 
is a normalized utility function that maps each coalition $S \subseteq P$ to a value in $[0,1]$.  
The goal is to approximate the Shapley value $\phi(p_j)$ of participant $p_j \in P$ with additive error $\varepsilon$. The algorithm employs three quantum registers: \textit{Partition register} $Q_{\mathrm{pt}}$ with $\ell$ qubits, used to encode an amplitude distribution proportional to the coalition-weight coefficients, $\omega(n, r)$, where $r = |S|$ and:
\begin{equation}    
\ell = \mathcal{O}\Big(\log\frac{(U_{\max}-U_{\min})\,n}{\varepsilon}\Big);
\end{equation}

The corresponding bounds for the normalized utility function $U$ are defined as
\begin{equation}
U_{\max} = \max_{S \subseteq P} U(S), 
\qquad 
U_{\min} = \min_{S \subseteq P} U(S).
\end{equation}

\textit{Player register} $Q_{\mathrm{pl}}$ with $n$ qubits, stores a superposition of all coalitions $S \subseteq P \setminus \{p_j\}$; and, the \textit{Utility register} $Q_{\mathrm{ut}}$ with a single qubit, represents the normalized utility of each coalition. Controlled rotations, $R_j$, parameterized by the partition amplitudes, prepare the superposition
\begin{equation}
\sum_{S \subseteq P \setminus \{p_j\}} \sqrt{\omega(n, |S|)}\,|S\rangle_{Q_{\mathrm{pl}}}, 
\end{equation}
so that, the amplitude of each coalition corresponds to its Shapley weight.  Two quantum oracles, $U_{\mathrm{val}}^{(+)}$ and $U_{\mathrm{val}}^{(-)}$, implement the normalized utility function $U$, 
conditioning on whether $p_j$ is included $(+)$ or excluded $(-)$ from the coalition. Applying the quantum amplitude estimation routine described in~\citet{qmcts} to these states yields the quantities $\phi^{(+)}(p_j)$ and $\phi^{(-)}(p_j)$, which correspond to the expected marginal contributions of $p_j$, when it is included in, and excluded from a coalition, respectively. The Shapley value is then obtained as
\begin{equation}
\phi(p_j) = \phi^{(+)}(p_j) - \phi^{(-)}(p_j)
\end{equation}
with total error bounded by $\varepsilon$ and overall query complexity
\(\mathcal{O}\Big(\frac{U_{\max}-U_{\min}}{\varepsilon}\Big)\), providing a near-quadratic speedup compared to classical Monte Carlo estimation. The speedup is achieved because quantum amplitude estimation requires only \(\mathcal{O}(1/\epsilon) \) queries to reach an additive error \( \epsilon \), compared to \(\mathcal{O}(1/\epsilon^2) \) queries for classical Monte Carlo approaches, thus substantially reducing computational costs. Formal proofs of correctness, error bounds, and complexity guarantees for the proposed quantum procedures are presented in~\citet{qmcts,qsv}.

%% file: sections/3_methodology.tex
\section{\QGShap}

\begin{figure}[ht]
    \centering
    \includegraphics[width=0.92\textwidth]{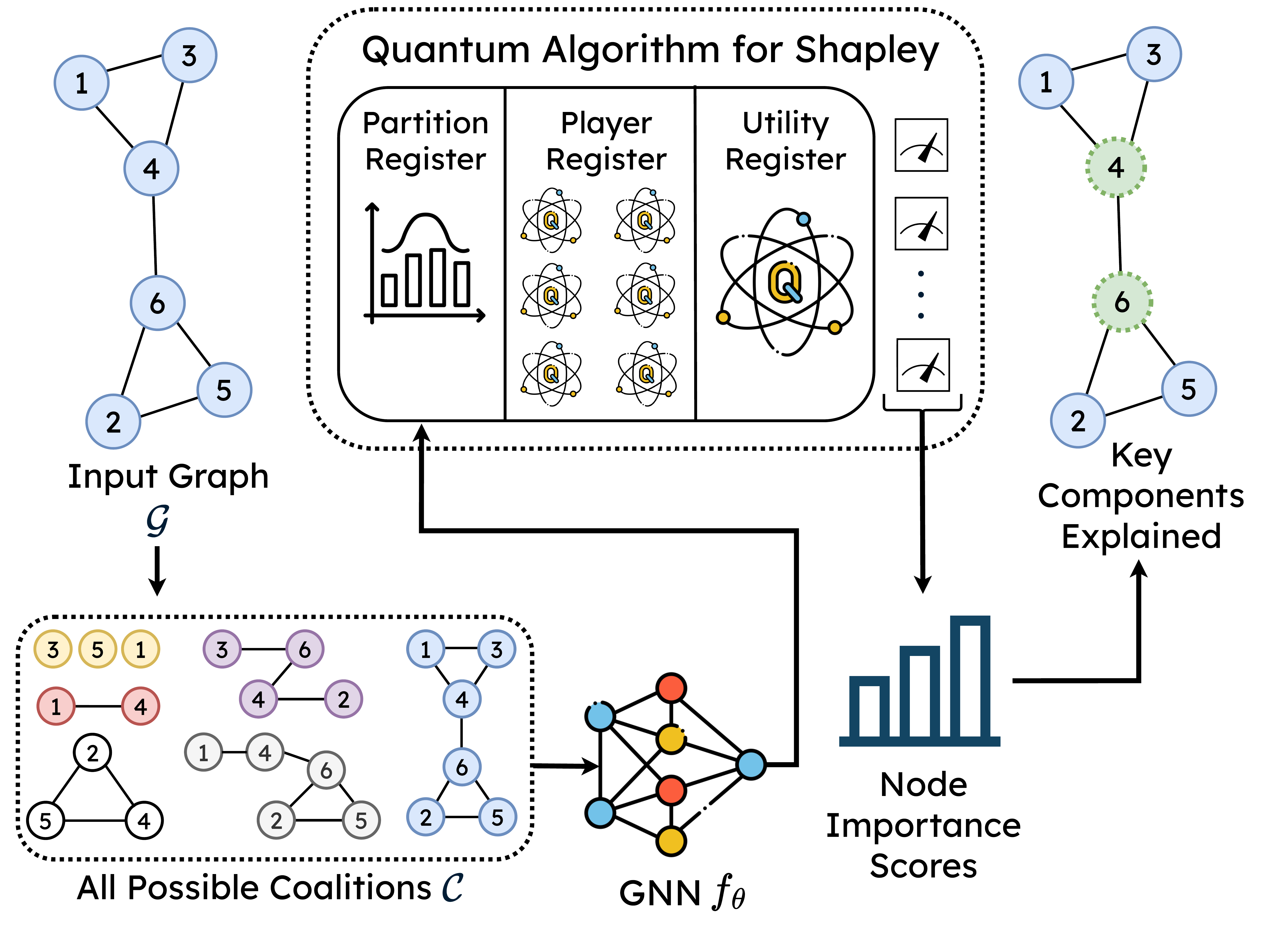}
     \caption{\QGShap Workflow: The input graph \(\mathcal{G}\) is mapped to node coalitions \(\mathcal{C}\), each scored by the trained GNN \(f_{\theta}\) via a masking oracle to obtain coalition values. A quantum module prepares three registers: a \emph{Partition} register encoding Shapley weights, a \emph{Player} register encoding coalition indices, and a \emph{Utility} register encoding normalized coalition scores. Quantum amplitude estimation over the \emph{Utility} register aggregates weighted marginal contributions, yielding node-level Shapley attributions as the final explanations.}
    \label{fig:qshap}
\end{figure}

We present \QGShap, a post hoc explainability framework that leverages quantum speedup to obtain exact Shapley value explanations for GNN predictions. Starting from a trained GNN, $f_{\theta}$, and an input graph, $G = (V, E)$, we exhaustively enumerate all non-empty node subsets, $S \subseteq V$, and generate corresponding masked graphs, $G_S$, via zero-fill encoding, where excluded nodes are replaced with zero vectors. Each masked graph is evaluated by $f_{\theta}$ to compute the cooperative game-theoretic value $v(S)$, ensuring that every node's marginal contribution is precisely captured without resorting to sampling heuristics.

Building on the quantum Shapley value estimation framework introduced in Section~\ref{sec:quantum_shapley} and originally proposed in~\citep{qsv} (Section~5), we adopt the same state-preparation procedure for encoding exact Shapley weights. For the amplitude estimation step, we employ the quantum routine described in~\citep{qmcts}. Accordingly, to prepare the exact Shapley weights in a quantum state, we first normalize the cooperative values:

\begin{equation}
\hat{v}(S) = \frac{v(S) - \min_{S'} v(S')}{\max_{S'} v(S') - \min_{S'} v(S')} \in [0,1].
\end{equation}

\begin{figure}[t]
    \centering
    \includegraphics[width=1.0\textwidth]{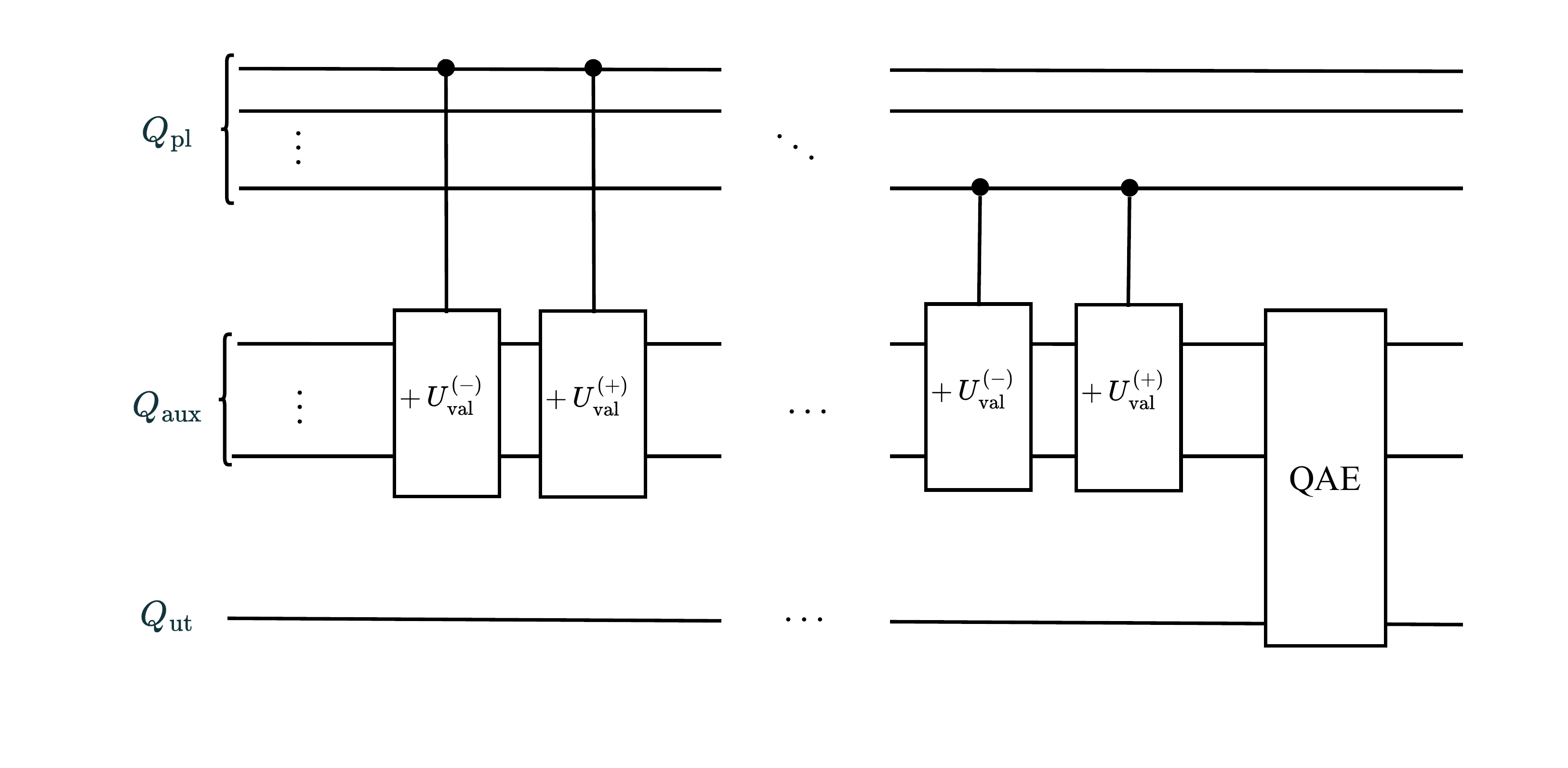}
    \caption{Circuit of the \QGShap\ utility oracle $U^{(\pm)}_{\text{val}}$. 
The player register $Q_{\text{pl}}$ encodes coalitions $S \subseteq V \setminus \{p_j\}$, while the auxiliary register $Q_{\text{aux}}$ and utility register $Q_{\text{ut}}$ store normalized cooperative values $\hat v(S)$ and $\hat v(S \cup \{p_j\})$. 
The quantum oracles $U^{(-)}_{\text{val}}$ and $U^{(+)}_{\text{val}}$ correspond to evaluating coalitions without and with node $p_j$, respectively. 
Quantum Amplitude Estimation (QAE) is then applied to $Q_{\text{ut}}$ to obtain the expected contributions $\phi^{(+)}(p_j)$ and $\phi^{(-)}(p_j)$, which are combined to reconstruct the exact Shapley value $\phi(p_j)$ as described in Section~\ref{sec:quantum_shapley}.}

    \label{fig:qshap_circuit_diagrams}
\end{figure}

We then allocate a \textit{player register} $Q_{\mathrm{pl}}$ of $|V|$ qubits, encoding coalition membership, a \textit{partition register} $Q_{\mathrm{pt}}$ initialized via beta-function rotations to load amplitudes proportional to the Shapley coefficients $w_{|S|,|V|}$, and a \textit{utility register} $Q_{\mathrm{ut}}$ to store the normalized cooperative value $\hat{v}(S)$ for each coalition. Controlled rotations between the partition and player registers prepare a superposition in which the amplitude of each basis state $\lvert S\rangle$ is proportional to $\sqrt{w_{|S|,|V|}}$. We then invoke the quantum amplitude estimation subroutine to extract the weighted expected contributions ${\phi}^{(+)}(p_j)$ and ${\phi}^{(-)}(p_j)$ from the utility register for each participant node $p_j$,
 achieving a quadratic reduction in sampling complexity compared to classical Monte Carlo methods. Finally, we reconstruct the Shapley value of each node by computing the difference in weighted expected values and denormalizing:


\begin{equation}
        \phi(p_j) = \bigl(\max_{S}v(S)-\min_{S}v(S)\bigr)\,\bigl(\phi^{(+)}(p_j)-\phi^{(-)}(p_j)\bigr).
\end{equation}

Then, we normalize again across all nodes to produce a global importance ranking. Thus, \QGShap computes Shapley contributions per node within each coalition, treating nodes as ``players'' in many small cooperative games, rather than treating entire coalitions or subgraphs as atomic units. For every coalition generated via exhaustive enumeration, it constructs all subsets of nodes within that coalition, evaluates the model's value function for each subset, and applies quantum amplitude estimation to compute the exact marginal Shapley value for each individual node in that coalition. This hierarchical, node-centric approach differs fundamentally from SubgraphX, which treats sampled subgraphs and residual nodes as players, simultaneously, as well as, GraphSHAP-IQ, which derives exact any-order Shapley interactions across nodes within receptive fields, but without explicit coalition traversal. The quantum speedup from amplitude estimation~\citep{qmcts} over normalized subset values within coalitions yields unbiased, high-fidelity node attributions in $\mathcal{O}(1/\epsilon)$ complexity, surpassing classical Monte Carlo's $\mathcal{O}(1/\epsilon^2)$, and motivates its application for precise quantum GNN explainability by capturing subtle intra-coalition interactions overlooked by SubgraphX's subgraph scoring and GraphSHAP-IQ's receptive-field constraints.

\QGShap's pipeline is illustrated in Fig.~\ref{fig:qshap}, the circuit of the quantum routine in Fig.~\ref{fig:qshap_circuit_diagrams}, and we formally detail \QGShap in Algorithm~\ref{alg:qshapleygnn}.

\begin{algorithm}[H]
\caption{\QGShap: Exact Shapley Value Estimation}
\label{alg:qshapleygnn}
\begin{algorithmic}[1]
\Input{Trained GNN $f_{\theta}$; input graph $G = (V, E)$}
\Output{Node-level Shapley value explanations $\{\phi_i\}_{i \in V}$}
\State Enumerate all non-empty subsets $\mathcal{C} = \{S \subseteq V : S \neq \emptyset\}$
\ForAll{$S \in \mathcal{C}$}
    \State Construct masked graph $G_S$ via zero-fill encoding
    \State Evaluate cooperative value $v(S) = f_{\theta}(G_S)$
\EndFor
\State Compute $v_{\min} = \min_{S \in \mathcal{C}} v(S)$ and $v_{\max} = \max_{S \in \mathcal{C}} v(S)$
\State Normalize all cooperative values: $\hat{v}(S) = \frac{v(S) - v_{\min}}{v_{\max} - v_{\min}}$ for all $S \in \mathcal{C}$
\State Allocate player register $Q_{\mathrm{pl}}$ ($|V|$ qubits), partition register $Q_{\mathrm{pt}}$, and utility register $Q_{\mathrm{ut}}$
\State Encode Shapley weights $w_{|S|,|V|}$ via beta-rotation circuits in $Q_{\mathrm{pt}}$
\State Apply controlled rotations $R_j$ to prepare superposition of coalitions in $Q_{\mathrm{pl}}$ and store $\hat{v}(S)$ in $Q_{\mathrm{ut}}$


\ForAll{$p_j \in V$}
    \State Apply quantum amplitude estimation on $Q_{\mathrm{ut}}$ to obtain $\phi^{(+)}(p_j), \phi^{(-)}(p_j)$
    \State $\phi(p_j) \gets (v_{\max} - v_{\min}) \cdot \bigl(\phi^{(+)}(p_j) - \phi^{(-)}(p_j)\bigr)$
\EndFor
\State Normalize $\{\phi(p_j)\}$ to produce final node importance scores

\end{algorithmic}
\end{algorithm}

%% file: sections/4_experiments.tex
\section{Experiments}
To thoroughly evaluate our proposed approach, we construct a controlled experimental setup that supports both quantitative and qualitative analysis of explanation performance. This section outlines the datasets, details the model used for prediction tasks, and describes the explanation methodology along with evaluation metrics applied to assess explanation effectiveness. Our implementation is available at \url{https://github.com/smlab-niser/qgshap}.

\subsection{Datasets}
\textbf{Bridge:} The Bridge detection dataset~\citep{stglime} consists of synthetically generated graphs formed by connecting two cycle graphs (3-5 nodes each) via a bridge linking selected nodes. Node identities are randomized across samples to ensure the model and explanations rely on graph structure rather than fixed node positions. The training set includes configurations with up to 15 nodes, containing both graphs with a bridge edge (label 1) and disconnected cycle pairs (label 0), totaling 60 balanced samples. The test set comprises 20 graphs with bridge edges, with exactly 8 nodes across four fixed configurations: (3+3), (3+4), (4+3), and (4+4), with each number denoting the order of the cycles connected with a bridge, enabling evaluation of generalization to unseen structures.\\

\noindent\textbf{BA2-Motif:} The BA2-Motif dataset~\citep{gnnexp} extends the Barab\'asi-Albert (BA) model into which exactly one motif is inserted either a house (label 1) or a cycle (label 0).  This dataset introduces motif-level explanation within scale-free topologies. Following the \textit{ExplainerDataset} setup in PyTorch Geometric, 50 train and 50 test graphs are generated. Notably, the house motif differs from a 5-cycle graph by including one additional edge closing the structure, designated as the `house edge'. The ability of an explainer to accurately pinpoint and explain the house edge becomes an explicit test for evaluating motif-level ground truth recovery in graph explanations.

\subsection{Model Training}
We implement a Graph Isomorphism Network (GIN)~\citep{gin} classifier using PyTorch 2.8.0\footnote{\url{https://pytorch.org/}} and PyTorch Geometric 2.6.1\footnote{\url{https://pytorch-geometric.readthedocs.io/en/2.6.1/}} (CUDA 12.8). The model consists of a multi-layer perceptron (MLP) as an encoder layer, three GIN layers, followed by another MLP as a decoder. We set the hidden dimension to 128. Training is conducted for 100 epochs using the Adam optimizer~\citep{adam} with a learning rate of $10^{-3}$. The model is optimized using binary cross-entropy loss between the predicted and true class labels. All test graphs in this study were limited to 8 nodes to ensure feasibility. In the quantum setting, each node requires a qubit in the player register, and the partition register must also scale with the number of nodes to encode coalition-weight amplitudes. Beyond 8 nodes, the number of qubits and circuit complexity grow rapidly, making exact Shapley value estimation infeasible with current quantum simulation resources. We employ the publicly available reference implementation\footnote{\url{https://github.com/iain-burge/QuantumShapleyValueAlgorithm}} to compute quantum-accelerated Shapley values for our experiments, executing all quantum subroutines on a Qiskit-based simulator\footnote{\url{https://www.ibm.com/quantum/qiskit}}.

\subsection{Evaluation Metrics}

We evaluate the quality of explanations using a set of standard metrics commonly used in the GNN explainability literature~\citep{gnnexp}, each capturing different aspects of explanatory performance.

\paragraph{\bfa{Top-$k$ Accuracy}}  
This metric measures the frequency with which the ground-truth target nodes appear among the top-$k$ most important nodes, as ranked by the explanation model. Formally, it is defined as:
\[
\mathrm{Acc}_{\mathrm{top-}k} = \frac{1}{N} \sum_{i=1}^N \mathbb{I}\left[\mathrm{target}_i \in \mathrm{Top}\text{-}k({\phi}(p_j))\right],
\]
where, ${\phi}(p_j)$ represents the importance scores assigned to nodes in instance $p_j$ (Shapley values for Shapley-based explainers, or standard node importance scores otherwise), and $\mathbb{I}[\cdot]$ is the indicator function. Only top-2 accuracy ($k=2$) is reported, since for both Bridge and BA2-Motif datasets, we focus on whether the explainer identifies either the bridge edge or the `house edge' that completes the house motif from a 5-cycle. Higher values indicate better alignment between the explainer's output and the true important nodes.

\paragraph{\bfa{Fidelity}}  
Fidelity~\citep{gnnexp} evaluates how well the explanation aligns with the model's predictions when key nodes are selectively retained or removed. 

\ul{
    \1 \ita{Fidelity-plus} ($\fidp$) measures the model's confidence or prediction consistency, when only the top-$k$ important nodes (by $\phi_i$) are retained, while all others are removed. 
    \1 \ita{Fidelity-minus} ($\fidm$) measures the effect of removing the top-$k$ nodes while keeping the rest.
}

 Formally, if \(G\) is the original graph, \(S\) is the set of top-\(k\) important nodes, and \(y_c\) is the predicted class on \(G\), then \(\fidp\) is defined as
\[
\fidp = P_{\text{keep}}(y_c) - P_{\text{base}}(y_c),
\]
where \(P_{\text{base}}(y_c)\) is the model's predicted class probability on the full graph \(G\), and \(P_{\text{keep}}(y_c)\) is the probability on the induced subgraph containing only \(S\). Conversely, 
\[
\fidm = P_{\text{base}}(y_c) - P_{\text{remove}}(y_c),
\]
where \(P_{\text{remove}}(y_c)\) is the predicted probability for class \(y_c\) on the complement graph \(G \setminus S\).

An effective explainer should yield high \( \fidp \) (the top-ranked nodes alone suffice to reproduce the model's prediction) and low \( \fidm \) (removing these nodes substantially changes the prediction), reflecting high confidence that it has correctly identified the most influential nodes driving the model's decision.

\paragraph{\bfa{Sparsity}}  
Sparsity~\citep{gnnexp} captures how concise the explanation is by computing the proportion of nodes that receive low importance scores:
\[
    S = 1 - \frac{|\{p_j : \phi(p_j) \ge 0.1 \max_i \phi(p_i)\}|}{N}.
\]
Here, $\phi(p_j)$ denotes the importance score of node $p_j$ (Shapley value for Shapley-based explainers, or the explainer's node importance score otherwise), and $N$ is the total number of nodes. Higher sparsity indicates that the explanation focuses on a small set of high-importance nodes, thereby improving explainability and reducing noise.

\paragraph{\bfa{Graph Explanation Accuracy (GEA)}} GEA~\citep{xgnnrev} quantifies how closely an explainer's predicted important nodes match the ground truth nodes in a graph, using the Jaccard similarity index. It is defined as:
\[
\mathrm{GEA} = \frac{\mathrm{TP}}{\mathrm{TP} + \mathrm{FP} + \mathrm{FN}}
\]
where, \(\mathrm{TP}\) (true positives) is the number of nodes correctly identified as important, \(\mathrm{FP}\) (false positives) is the number of nodes wrongly identified as important, and \(\mathrm{FN}\) (false negatives) is the number of true important nodes missed. GEA yields a value between 0 (no overlap) and 1 (perfect match), offering an intuitive, symmetric measure of explanatory set quality by considering both types of errors equally. Moreover, unlike $\fidp$ or $\fidm$, which can remain high even when an explainer assigns maximal score to the wrong nodes due to distributional sufficiency effects, GEA directly penalizes such target misalignments by measuring set overlap with ground truth, thereby detecting cases where high fidelity coexists with incorrect node attributions.

%% file: sections/5_results.tex
\subsection{Results and Discussion}

\begin{figure}[ht]
    \centering
    \begin{tabular}{cc}
        \includegraphics[width=0.30\textwidth]{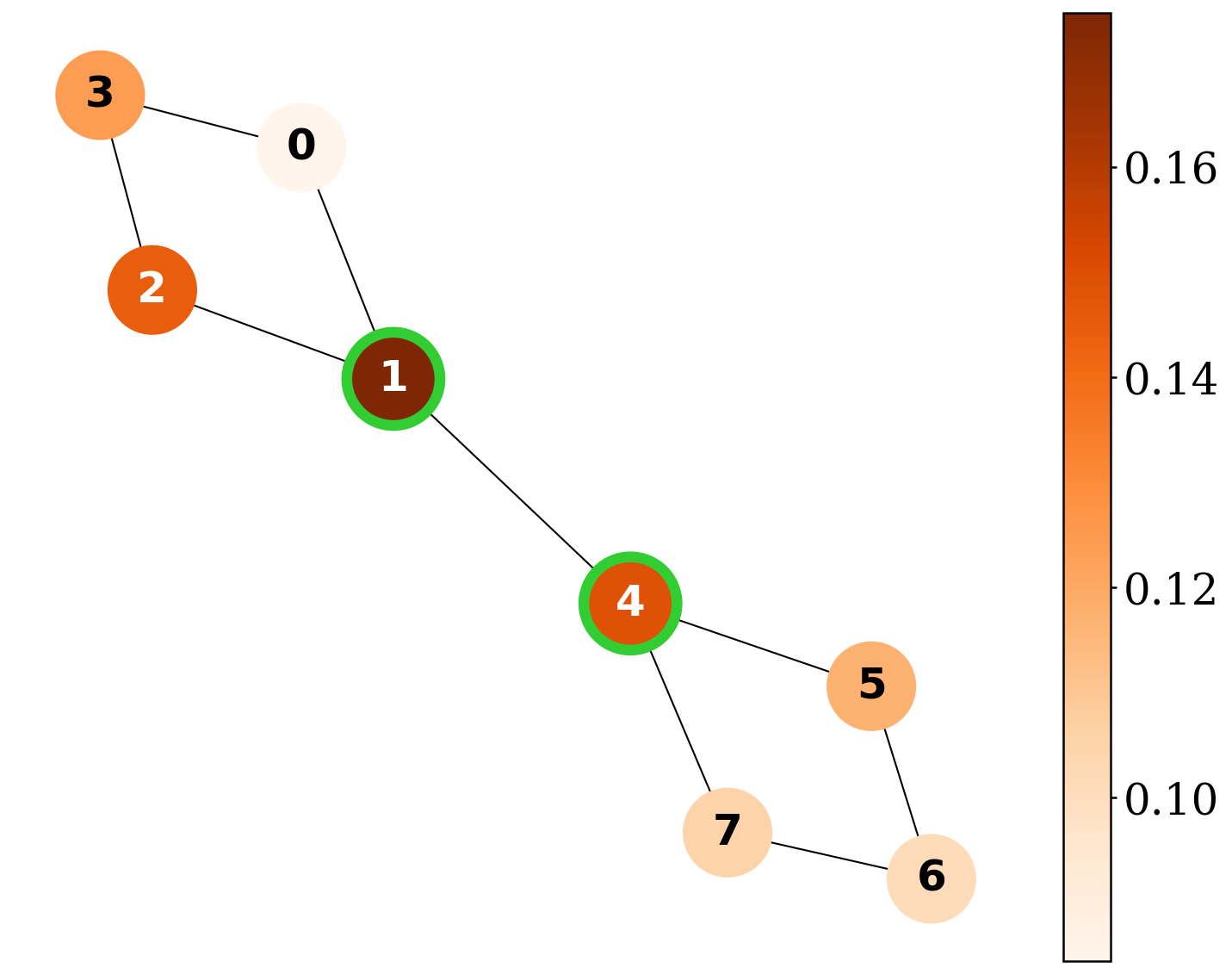} &
        \includegraphics[width=0.30\textwidth]{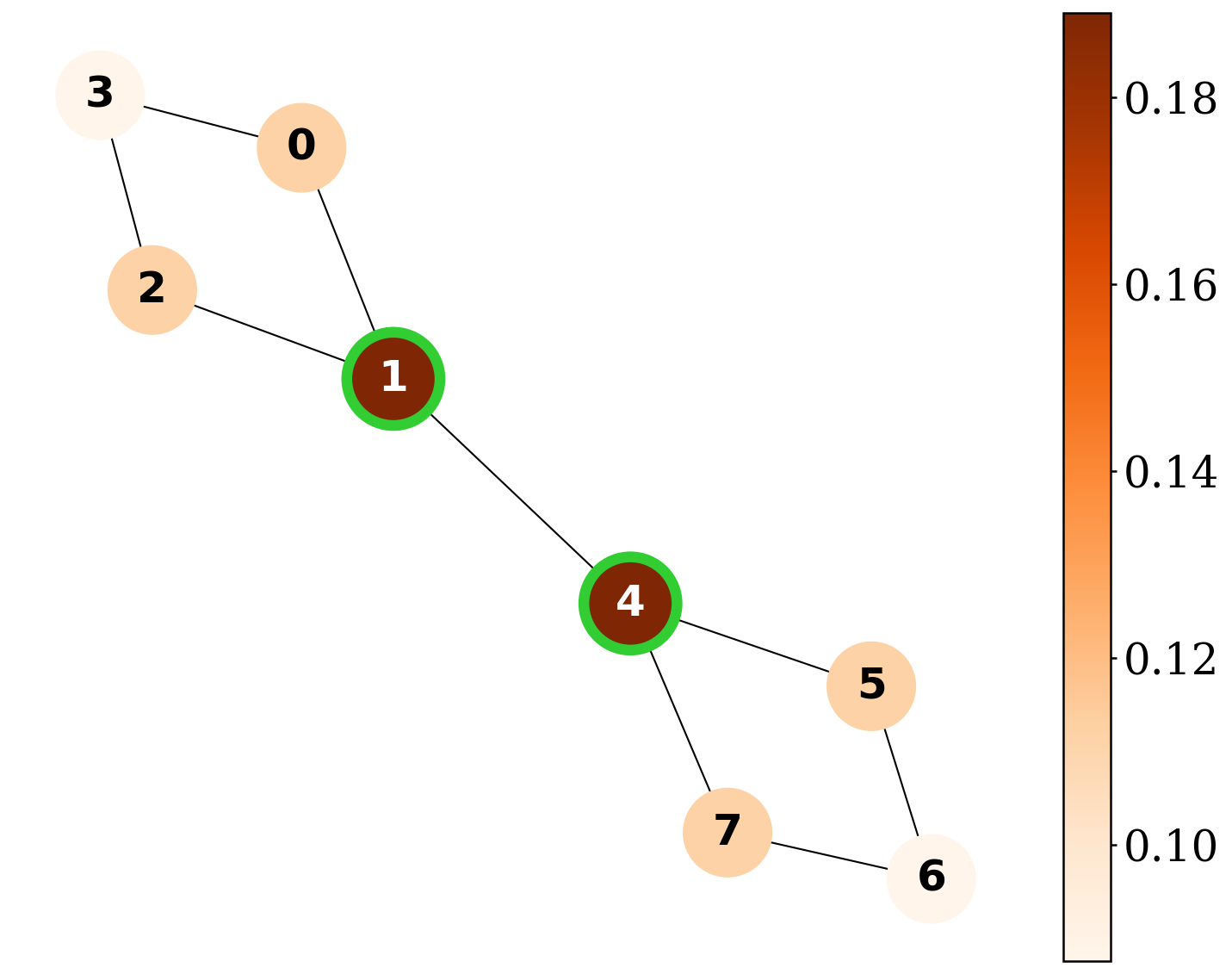} \\
        \includegraphics[width=0.30\textwidth]{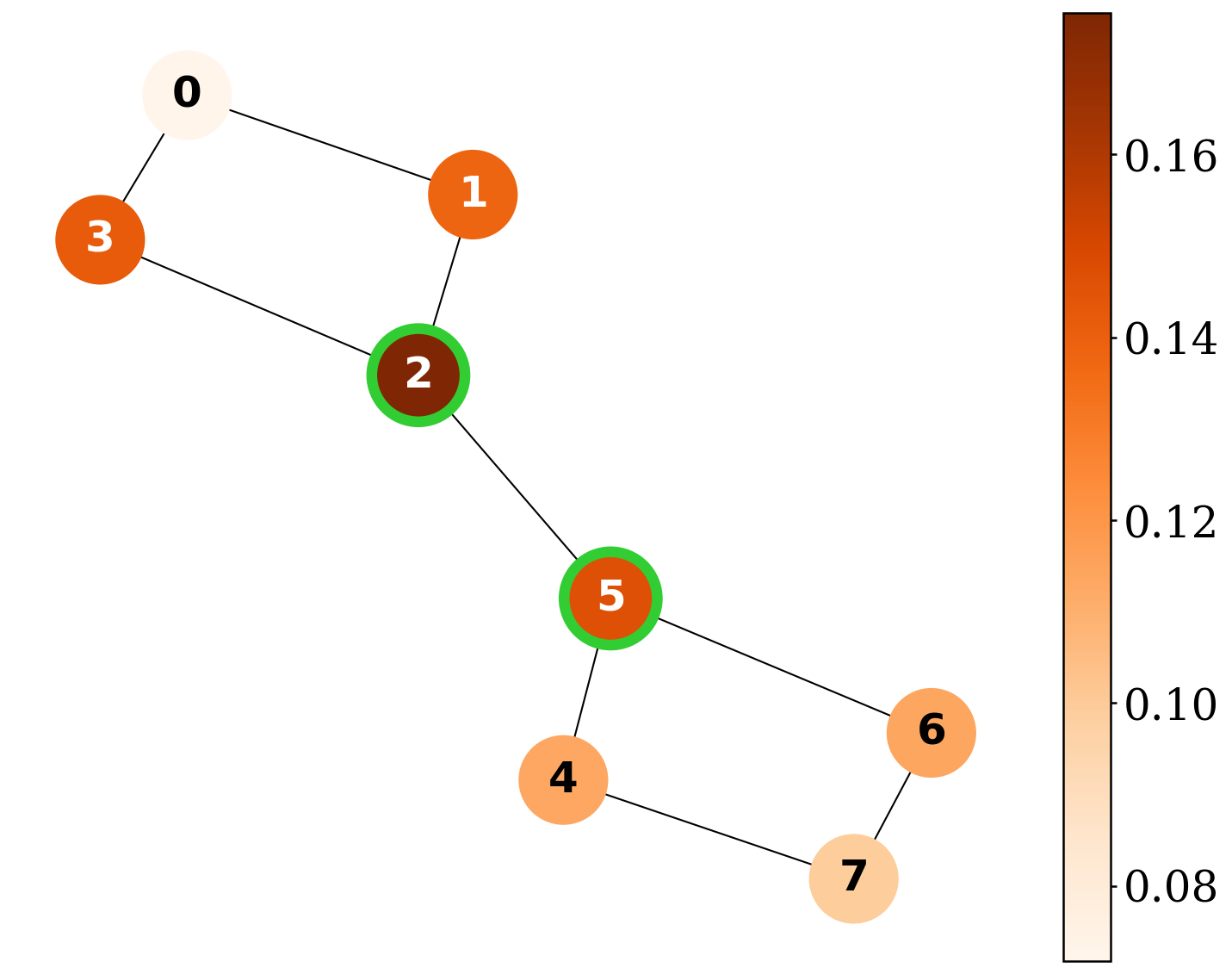} &
        \includegraphics[width=0.30\textwidth]{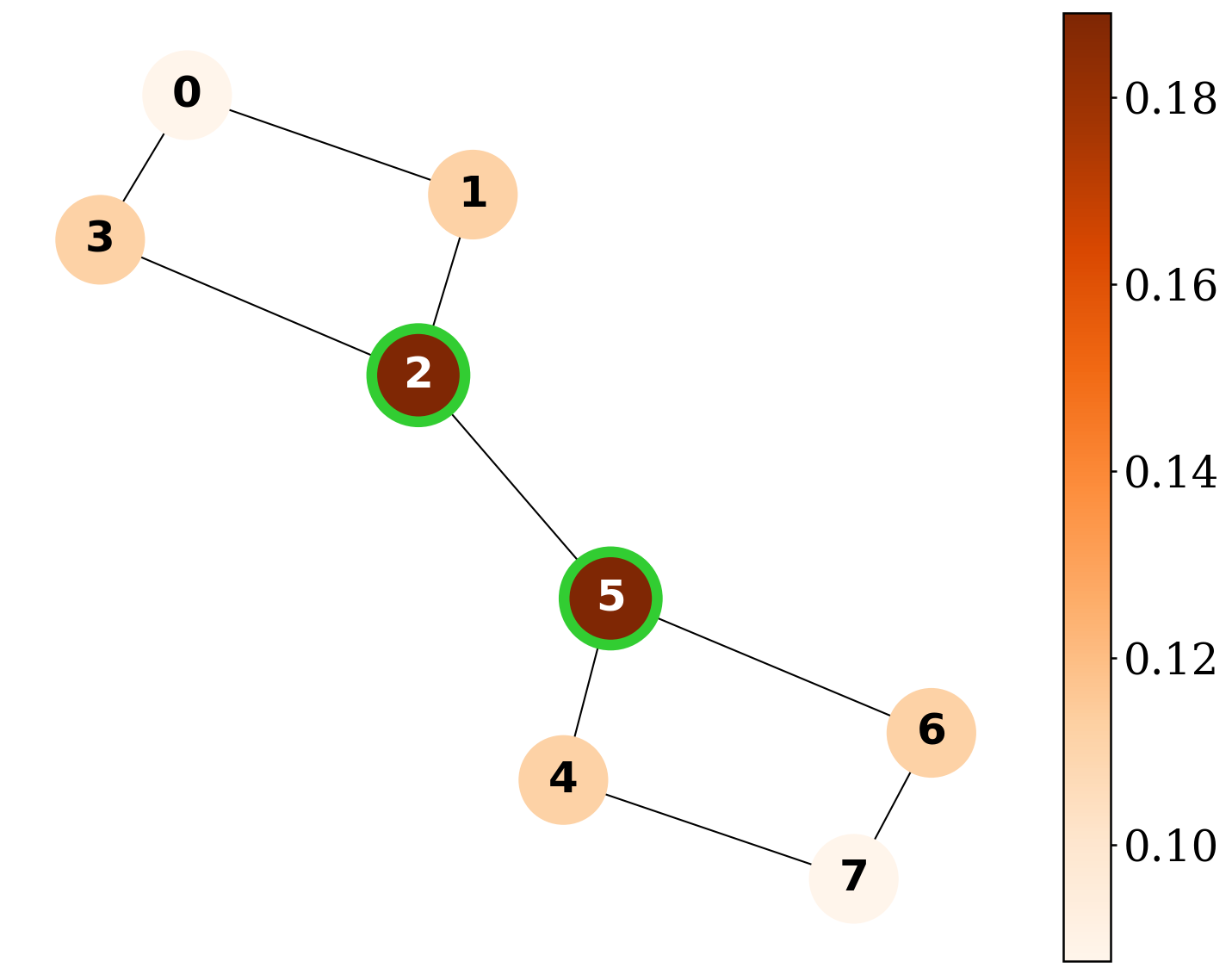} \\
    \end{tabular}
    \caption{\bfa{Bridge}: Subgraph explanations using SubgraphX and \QGShap. \textbf{Left - }SubgraphX . \textbf{Right - }\QGShap}
    \label{fig:brres}
\end{figure}

\begin{figure}[ht]
    \centering
    \begin{tabular}{cc}\includegraphics[width=0.30\textwidth]{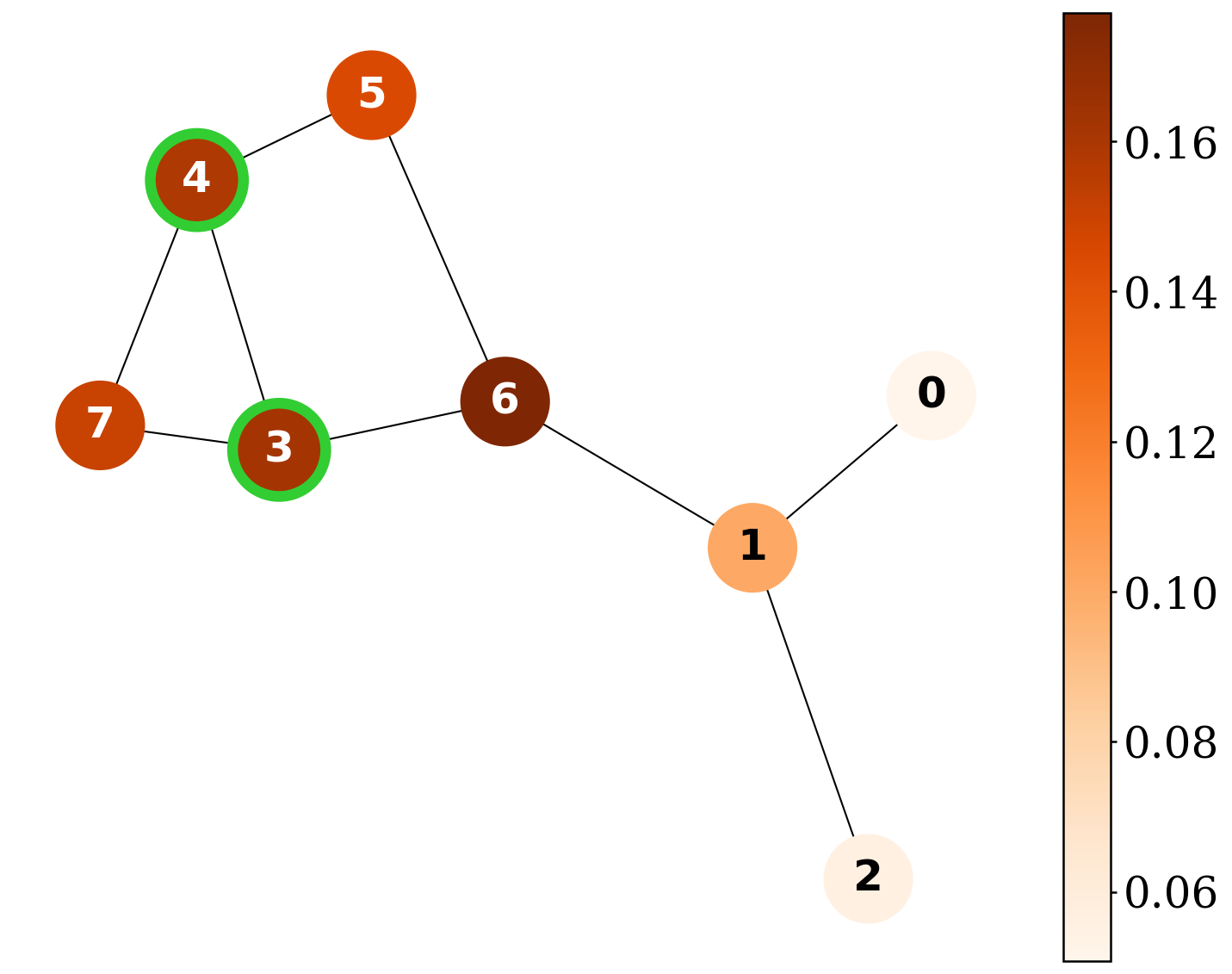} &
        \includegraphics[width=0.30\textwidth]{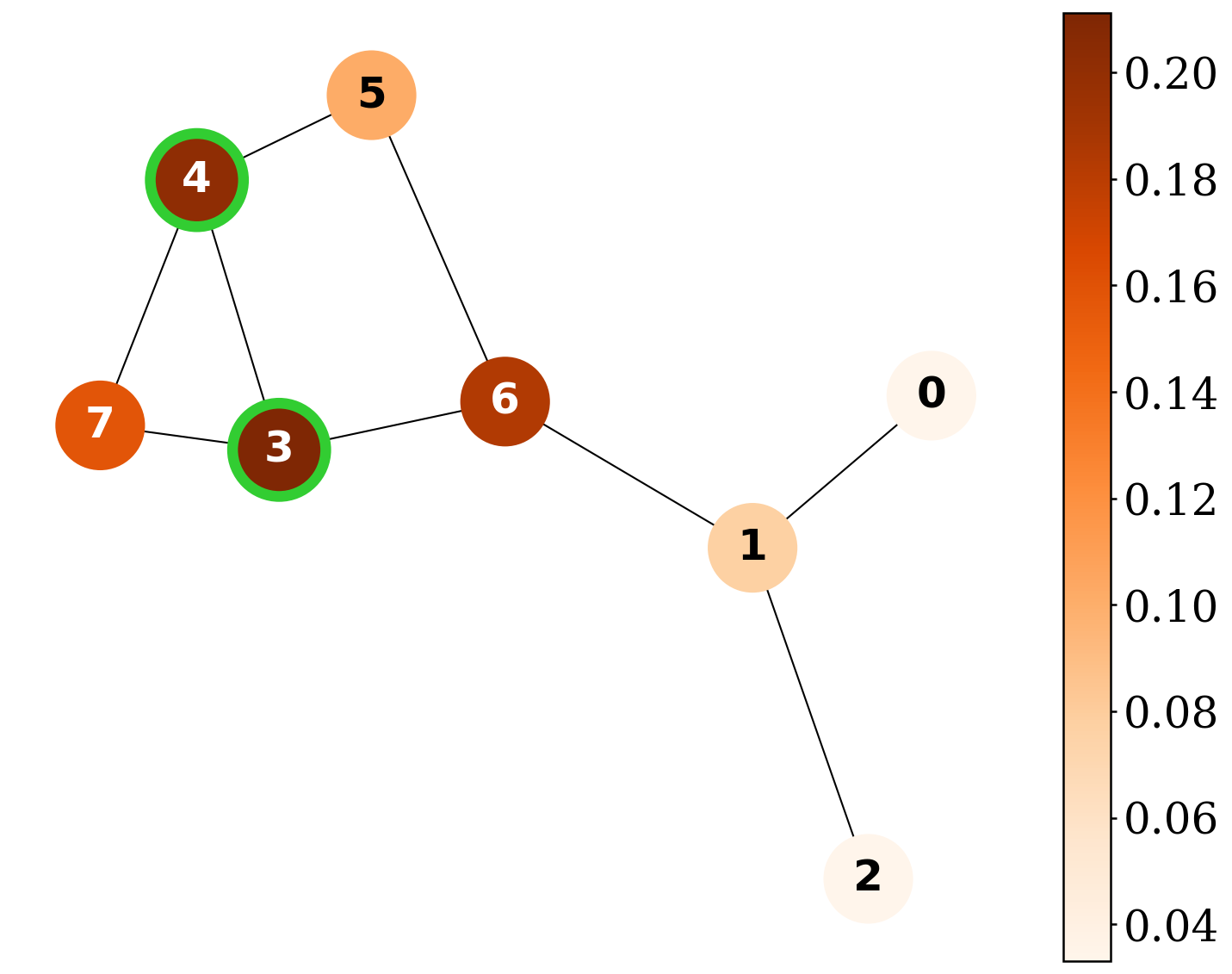} \\
        \includegraphics[width=0.30\textwidth]{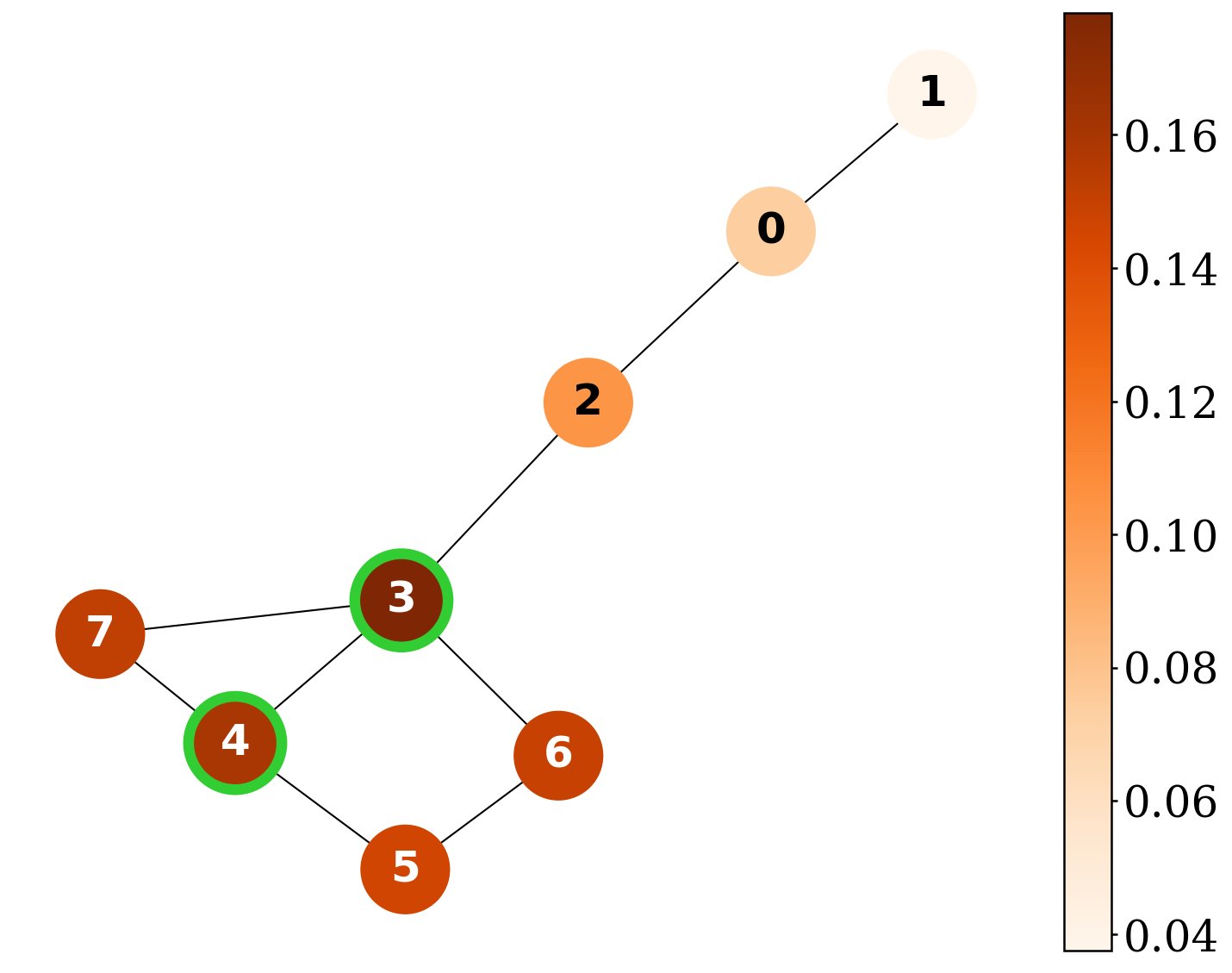} &
        \includegraphics[width=0.30\textwidth]{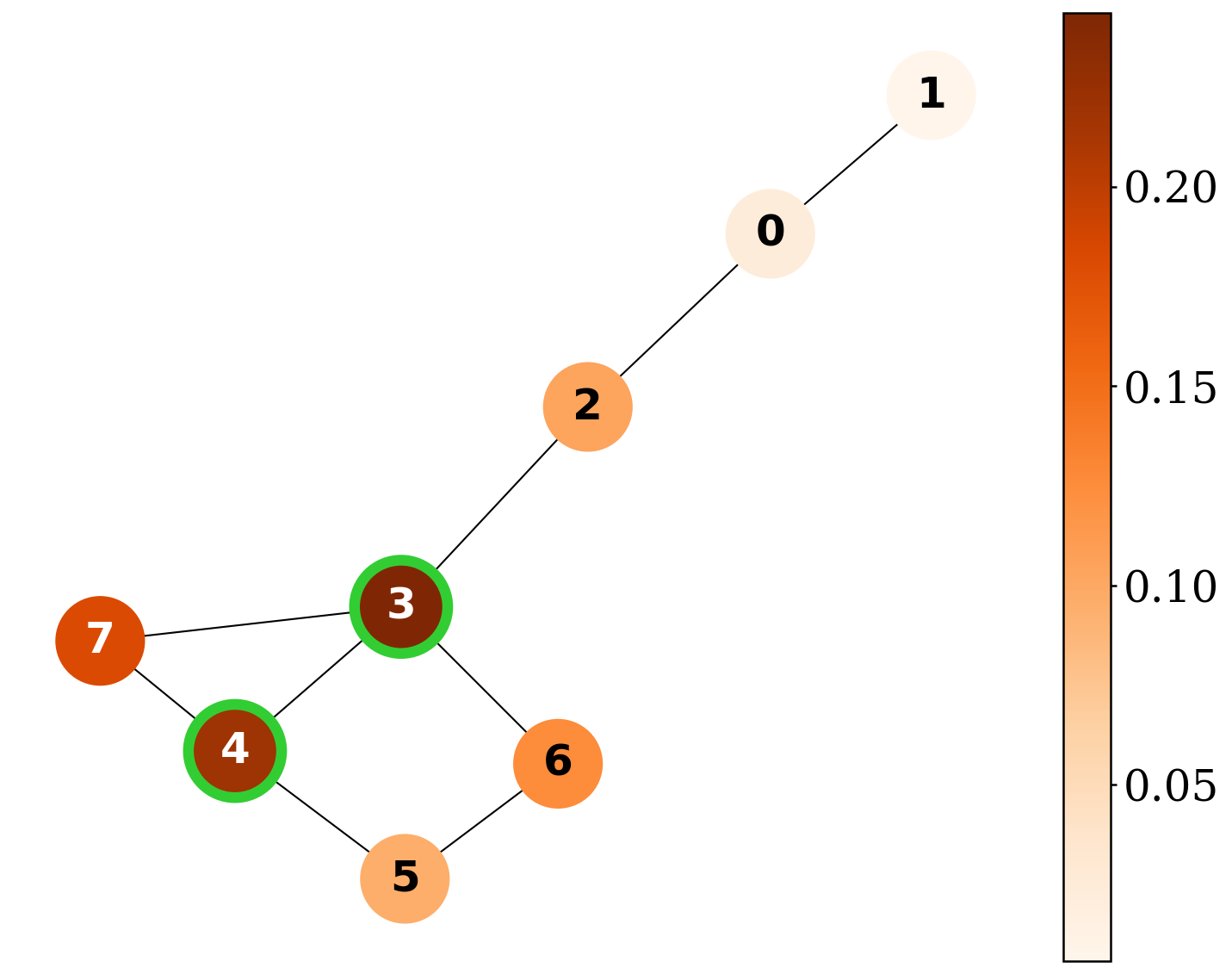} \\
    \end{tabular}
    \caption{\bfa{BA2-Motif}: Subgraph explanations using SubgraphX and \QGShap. \textbf{Left - } SubgraphX . \textbf{Right - } \QGShap}
    \label{fig:bares}
\end{figure}

Visualizations for the Bridge and BA2-Motif datasets are provided in Fig.~\ref{fig:brres} and~\ref{fig:bares}, with quantitative metrics summarized in Table~\ref{tab:bridge_ba_metrics}. We show graph heatmaps only for SubGraphX, as it is the sole competitive Shapley-based explainer among the compared methods.

\begin{table}[hb]
\caption{Comparison of explanation metrics for different explainers.}
\centering

\begin{adjustbox}{max width=\linewidth}
\begin{tabular}{l l 
                r@{\hspace{6pt}} 
                r@{\hspace{6pt}} 
                r@{\hspace{6pt}} 
                r@{\hspace{6pt}} 
                r}
\toprule
\textbf{Dataset} & \textbf{Explainer} 
& \multicolumn{1}{c}{\textbf{$\fidp$}} 
& \multicolumn{1}{c}{\textbf{$\fidm$}} 
& \multicolumn{1}{c}{\textbf{Sparsity}} 
& \multicolumn{1}{c}{\textbf{GEA}} 
& \multicolumn{1}{c}{\textbf{Top-2 Acc}} \\
\midrule
\multirow{4}{*}{Bridge} 
& GNNExplainer  & $0.60 \pm 0.49$ & $1.00 \pm 0.00$ & $0.75 \pm 0.00$ & $\sbest{0.07 \pm 0.13}$ & $\sbest{0.10 \pm 0.20}$ \\
& PGExplainer   & $\sbest{0.99 \pm 0.00}$ & $1.00 \pm 0.00$ & $0.75 \pm 0.00$ & $0.00 \pm 0.00$ & $0.00 \pm 0.00$ \\
& SubgraphX     & $\best{1.00 \pm 0.00}$ & $1.00 \pm 0.00$ & $0.75 \pm 0.00$ & $\best{1.00 \pm 0.00}$ & $\best{1.00 \pm 0.00}$ \\
& \textbf{\QGShap (Ours)} & $\best{1.00 \pm 0.00}$ & $1.00 \pm 0.00$ & $0.75 \pm 0.00$ & $\best{1.00 \pm 0.00}$ & $\best{1.00 \pm 0.00}$ \\
\midrule
\multirow{4}{*}{BA2-Motif} 
& GNNExplainer  & $\best{1.00 \pm 0.00}$ & $1.00 \pm 0.00$ & $0.75 \pm 0.00$ & $\sbest{0.32 \pm 0.11}$ & $0.50 \pm 0.19$ \\
& PGExplainer   & $0.01 \pm 0.01$ & $1.00 \pm 0.00$ & $0.75 \pm 0.00$ & $0.00 \pm 0.00$ & $0.00 \pm 0.00$ \\
& SubgraphX     & $\best{1.00 \pm 0.00}$ & $1.00 \pm 0.00$ & $0.75 \pm 0.00$ & $\best{0.40 \pm 0.00}$ & $\sbest{0.79 \pm 0.25}$ \\
& \textbf{\QGShap (Ours)} & $\best{1.00 \pm 0.00}$ & $1.00 \pm 0.00$ & $0.75 \pm 0.00$ & $\best{0.40 \pm 0.00}$ & $\best{1.00 \pm 0.00}$ \\
\bottomrule
\end{tabular}
\end{adjustbox}

\label{tab:bridge_ba_metrics}
\end{table}

\paragraph{\bfa{Bridge}}
On the Bridge dataset, our proposed \QGShap consistently achieved perfect performance across all evaluation metrics, reflecting both precision and reliability in its explanations. As shown in Table~\ref{tab:bridge_ba_metrics}, \QGShap attained Fidelity$^+$, GEA, and Top-2 Accuracy scores of $1.00 \pm 0.00$, matching or surpassing the strongest baseline, SubgraphX. In every case, it successfully identified the critical bridge nodes driving the model's predictions, demonstrating clear explainability and stability. These results highlight \QGShap's ability to deliver faithful and consistent explanations that align closely with the underlying graph structure and decision logic.  

\paragraph{\bfa{BA2-Motif}}
For the BA2-Motif dataset, \QGShap also produced highly competitive and insightful results, achieving the highest Top-2 Accuracy ($1.00 \pm 0.00$) among all explainers. In most instances, it correctly pinpointed the two key nodes and their connecting edge responsible for predicting the `house' motif, capturing the core structural reasoning of the model. Even in a few challenging cases, where not all motif nodes were ranked at the top, \QGShap consistently prioritized the most influential node pairs linked to the correct class. This behavior underscores its robustness and interpretive strength in highlighting the decisive substructures within complex graph motifs.

%% file: sections/6_conclusion.tex
\section{Conclusion}

We introduce \QGShap, a post hoc explainability framework that combines cooperative game theory and quantum computation to produce exact Shapley value explanations for GNN predictions. Unlike classical sampling or approximation-based methods, \QGShap evaluates all coalitions, capturing node influence precisely and verifiably rather than relying on heuristics. Although currently limited to small graphs by hardware constraints, \QGShap demonstrates that quantum computation can make exact, \emph{classically intractable} Shapley calculations practical and establishes a benchmark for evaluating classical explainers while bridging explainable GNNs with developments in quantum computing. Extending both the GNN and explanation modules into the quantum domain, our framework provides a principled and scalable approach to explainability in GNNs, classical or quantum. Future work could explore iterative, noise-resilient amplitude amplification strategies for robustness under realistic hardware constraints and fault-tolerant settings.

\subsection*{Limitations}
Although the method offers a near-quadratic speedup over classical Monte Carlo techniques up to polylogarithmic factors, \QGShap\ remains constrained to small graphs due to the exponential number of coalitions and gate preparations, which increases qubit requirements and circuit depth. The practical cost of classical simulation remains high: even on a system with a 48-core AMD CPU and an Nvidia A100 GPU (80\,GiB VRAM), the Bridge and BA2-Motif experiments required approximately 31 and 42 hours of runtime, respectively. Moreover, quantum noise and decoherence can reduce the precision of amplitude estimation, and access to high-quality quantum hardware remains limited. These factors underscore the current operational scope of \QGShap and highlight the need for advances that enable scaling to larger and more complex graphs.